# Wealth Taxes and Post-Growth: How different tax designs align with different goals

Thomas Webb*[1], Arthur Apostel[2], Milena Büchs[1], Richard Bärnthaler[1]

* corresponding author, ee22tcw@leeds.ac.uk

[1] Sustainability Research Institute, School of Earth, Environment and Sustainability, University of Leeds, Leeds, LS2 9JT, United Kingdom

[2] Faculty of Economics and Business Administration, Ghent University, Campus Tweekerken, Tweekerkenstraat 2, 9000 Ghent, Belgium

**Abstract**

Wealth taxes are a frequently proposed policy within the post-growth literature, but evaluations of their alignment with post-growth goals, and empirical estimates of their potential effects, are lacking. We contribute to this literature by examining the extent to which different wealth-tax designs can contribute to four goals of a post-growth transition: redistributing wealth; eradicating extreme wealth; curbing rent-seeking; and reducing $CO_2$ emissions. The analysis is based on microsimulation modelling, using household-level data from 18 countries of the 2017 EU Household Finance and Consumption Survey. Our analysis finds that taxes on net wealth are the most progressive and redistributive, while taxes on financial and investment property wealth tend to be more effective at addressing rent-seeking. However, we also identify trade-offs and conflicts between different tax designs and goals. As a result, a broader package of policies will be necessary to navigate these conflicts and mitigate the limitations inherent in any single wealth-tax design.

**Highlights**

- Wealth taxes can contribute to key goals of a post-growth transition: redistributing wealth, eradicating extreme wealth, curbing rent-seeking, and reducing $CO_2$ emission reduction.
- Taxes on net wealth are the most redistributive.
- Taxes on financial and investment property wealth are most effective at reducing rent-seeking.
- Trade-offs exist between different goals, necessitating policy packages.

# 1. Introduction

Humanity faces a polycrisis (Lawrence *et al.*, 2024) – a set of intertwined and mutually reinforcing crises – including climate and ecological breakdown as well as extreme, deepening inequality. Six of the nine planetary boundaries have now been transgressed (Sakschewski *et al.*, 2025), and wealth ownership is becoming ever-more concentrated (Piketty and Saez, 2014; Piketty and Zucman, 2014; European Environmental Bureau, 2024). Attempts to tackle these tendencies from within the growth paradigm appear increasingly inadequate, due to insufficient decoupling of GDP growth from environmental harm (Haberl *et al.*, 2020; Vadén *et al.*, 2020; Vogel and Hickel, 2023) and the failure of growth to reduce inequality (Hickel, 2017; Christensen *et al.*, 2023). As a result, post-growth – the science of well-being within planetary boundaries (Kallis *et al.*, 2025) – is exploring alternatives to the goal of continuous economic expansion. Instead, it focuses on policies that prioritise the enhancement of human well-being and social equity within planetary boundaries (O'Neill *et al.*, 2018), while also promoting a more emancipatory politics (Laruffa, 2025).

In the current growth-oriented capitalist political economy, social outcomes related to well-being are widely perceived to depend on ever-increasing GDP and productivity (Büchs and Koch, 2019). In response, sustainable welfare has emerged as a key topic within eco-social policy and post-growth research (Koch and Hansen, 2024), exploring how welfare states can contribute to universal and sustainable need satisfaction (Hirvilammi and Koch, 2020; Büchs *et al.*, 2024a). Welfare state funding, however, is typically framed as relying on income and consumption taxes and income-related social security contributions. The size of these revenue sources tends to be coupled to changes in economic output (Raphael *et al.*, 2024; Büchs *et al.*, 2024a), which would likely contract in a post-growth scenario in the Global North.

Post-growth welfare funding is therefore one of the 'bottleneck questions' in eco-social policy and sustainable welfare research (Bohnenberger, 2023). A range of fiscal and monetary proposals for a post-growth context has been explored (Jackson and Jackson, 2021; Olk *et al.*, 2023), and among these, taxes on wealth have received growing attention (Bailey, 2015; Hickel *et al.*, 2022; Koch, 2022). Because wealth represents a *stock* rather than a *flow*, some authors argue that it may be more resilient to economic fluctuations in GDP (Koch, 2022; Büchs *et al.*, 2024a). Another prominent argument for wealth taxation in the post-growth context is that inequality is likely to increase without growth and targeted intervention (Piketty and Saez, 2014; Jackson and Victor, 2016). Wealth taxes could therefore be an important measure for reducing inequality in a post-growth context (Murphy, 2023; Kallis *et al.*, 2025; Büchs *et al.*, 2024a).

Depending on underlying assumptions about the mechanisms of government spending, wealth taxes are understood to either directly raise revenue for welfare spending (e.g. Buch-Hansen and Koch, 2019; Fitzpatrick *et al.*, 2022; Engler *et al.*, 2024) or – following Modern Monetary Theory (MMT) – to operate primarily by removing money from circulation, thereby reducing aggregate demand and helping maintain price stability, which in turn creates macroeconomic space for public spending (Olk *et al.*, 2023). In both cases, wealth taxes play an important role, and both perspectives recognise their significance for distributive justice as well as for shaping incentives and behaviours that influence the allocation of real resources.

In addition, ecological and post-Keynesian macro-economists argue that wealth taxes can act as macroeconomic stabilisers during or after a post-growth transition. For example, Cahen-Fourot and Lavoie (2016) show that positive net profit and interest rates are compatible with a zero-growth economy *if* accumulated wealth is sufficiently recirculated – through consumption or other

expenditures – so that money flows back into the productive sectors and enables agents to service their debts; and Kemp-Benedict (2025) argues that wealth taxes can be used to achieve this.

Other justifications for taxing wealth also draw on the well-established 'ability-to-pay' and 'polluter-pays' principles: high-wealth households have greater capacities to bear fiscal burdens – including through the income and privileges their assets generate (Schmiel, 2024) – and their disproportionately large environmental footprints (see Section 2.1) make them responsible for covering a larger share of ecological mitigation costs (Büchs *et al.*, 2024b; Chancel *et al.*, 2025).

Although the post-growth literature frequently discusses wealth taxes, it has not yet evaluated how concrete wealth-tax designs align with specific post-growth goals. This study addresses this gap by examining the effects of different wealth-tax designs in relation to four post-growth goals:

1. wealth redistribution
2. eradication of extreme wealth
3. curbing rent-seeking
4. $CO_2$ emission reduction

Using data from the EU Household Finance and Consumption Survey (HFCS) and microsimulation modelling, we analyse the impacts of 12 wealth tax designs on these post-growth goals. We identify and examine trade-offs and conflicts between different goals and tax designs. The remainder of the article is structured as follows: Section 2 reviews the existing post-growth and empirical literature on wealth taxes and provides details on the four post-growth goals; Section 3 details the data and methods used; Section 4 presents our results; Section 5 contextualises the results, explores potential trade-offs between the tax designs, and discusses methodological limitations of this study; and Section 6 concludes.

## 2. Literature review

Recurrent taxes on net wealth were once more common in Europe. In 1990, twelve OECD countries levied taxes on net wealth, while today only three still do so (OECD, 2018). The widespread repeal of wealth taxes formed part of a broader trend of tax reductions that began in the 1980s (OECD, 2018). However, rising wealth concentration and perceived pressures on public finances (O'Donovan, 2021) have renewed calls for their reintroduction, both within and beyond post-growth debates. Section 2.1 summarises the existing empirical research on wealth taxes, and section 2.2 outlines how they could contribute to post-growth goals. The literature review focusses on net wealth taxes, primarily because the overwhelming majority of existing empirical studies do the same.

### 2.1 Empirical wealth tax research: effects on distribution, revenue, and emissions

There are surprisingly few academic studies that examine how a wealth tax affects the distribution of wealth. Apostel and O'Neill (2022) model the effects of different one-off wealth tax proposals in Belgium and find that the wealth share held by the top 1% could decrease by 0.16 to 0.84 percentage points (from 23.9%), and the wealth share held by the top 5% by 0.12 to 0.91 percentage points (from 42.2%), depending on the tax proposal and levels of avoidance. Marti *et al*. (2023) analyse differences in wealth inequality and wealth tax design across Swiss Cantons and find that a 0.1% reduction in top marginal annual wealth tax rates results in an increase of 0.9 percentage points in the wealth share of the top 1%, and 1.2 percentage points for the top 0.1%, five years after the reform. However, the authors note that misreporting of wealth and intra-country mobility are likely to inflate these results.

Since wealth is unevenly distributed along gender, generational, and geographical lines (Parkes and Johns, 2024; Advani *et al*., 2020c), a wealth tax would also help to address these inequalities.

Some authors have estimated the revenue that a wealth tax could generate across Europe. Apostel and O'Neill (2022) estimate that a wealth tax could generate between €5.9bn and €43.1bn in Belgium, depending on the tax design and levels of avoidance. Kapeller *et al.* (2023) model four different tax designs in Europe with varying degrees of progressivity, estimating revenues between €192bn and €1,281bn. Similarly, Krenek and Schratzenstaller (2018) simulate that a tax of 1% on wealth above €1m and 1.5% on wealth above €5m across 20 EU countries could yield approximately €156bn. Piketty (2014, p.528) estimates that a progressive Europe-wide tax with rates up to 2% on wealth over €5m could raise approximately €300bn. Finally, Landais *et al*. (2020) estimate that a progressive EU wealth tax with rates up to 3% for billionaires could raise 1.05% of EU GDP (or approximately €162bn in 2020) every year.

Regarding the environmental effects of wealth taxes, evidence remains scarce. Apostel and O'Neill (2022) use the wealth inequality-$CO_2$ emissions elasticity from Knight *et al*. (2017) to estimate that a one-off wealth tax in Belgium could have a positive but small effect – reducing emissions by between 0.1% and 0.6%, depending on tax rates. However, further research is needed to better understand the environmental implications of wealth taxes – in particular, their impact on investment behaviour and associated emissions.

## 2.2 How wealth taxes can contribute to the goals of a post-growth transition

Wealth taxes are widely regarded as a core component of post-growth policy (Fitzpatrick *et al*., 2022). However, scholars envision wealth taxes serving a range of purposes, which can be broadly categorised into four goals (in no particular order):

**Goal number 1: Redistribute wealth**. Achieving a more equitable distribution of wealth is often the main goal of wealth taxes in post-growth literature and is a near-universal aim across post-growth policy research (Fitzpatrick *et al*., 2022). Redistribution is seen as a worthwhile objective in itself, especially during a post-growth transition, as inequality may otherwise increase without additional predistributive, redistributive, or protective measures (Piketty and Saez, 2014; Jackson and Victor, 2016).

**Goal number 2: Eradicate extreme wealth.** Some authors take the goal of redistribution further, arguing for the need to eradicate extreme wealth altogether (e.g. Buch-Hansen and Koch, 2019). Those working within sufficiency or 'limitarian' frameworks (Robeyns, 2022) emphasise the importance of imposing ceilings on consumption, production, and affluence (Gough, 2022; Khan *et al*., 2023; Bärnthaler, 2024). Wealth taxes are one tool that can contribute to this objective. Recent work by the New Economics Foundation explores the idea of an 'extreme wealth line' (Balata *et al.*, 2025), arguing that such a threshold could help to identify the point at which extreme wealth becomes harmful to society, democracy, and the environment, and could help challenge the notion of limitless wealth accumulation. However, there is no consensus on what constitutes extreme wealth or on where such thresholds should be set (Buch-Hansen and Koch, 2019). In a survey of fifteen millionaires in six middle-to-high income countries, Balata *et al*., (2025) find that the most popular absolute threshold for 'extreme' wealth is $10m (equivalent to €8.9m in May 2025), while Robeyns *et al*. (2021) find that the Dutch public primarily place the threshold between €1m and €3m. Balata *et al*. (2025) also survey politicians and policy experts and report a preference for relative thresholds. In our analysis, we therefore use both an absolute threshold of €8.9m following the findings by Balata *et al*. (2025), and a relative threshold based on the 99$^{th}$ percentile of net wealth (approximately €2.4m – in line with Robeyns *et al.* (2021)), in order to explore two suitably distinct cut-offs.

**Goal number 3: Curb rent-extraction.** The third goal involves shifting the economy away from rent-extracting activities to improve needs satisfaction, reduce growth dependencies, and address a key process through which inequality is (re)produced (Stratford, 2020). Rent extraction takes many forms – for example, the for-profit provision of housing by landlords, and the growing financial sector, which increasingly "redistribute[s] money to [itself], at great cost to the rest of society" (Dietz and O'Neill, 2013, p.104) and extracts value created elsewhere (Mazzucato, 2018; Bärnthaler and Gough, 2023). Such rent extracting activities impede social-ecological transformation (Stratford, 2020) and should form as small a share of a post-growth economy as possible (Dietz and O'Neill, 2013, p.110). In this context, Fanning *et al*. (2020) introduce the concept of 'appropriating systems', which – unlike provisioning systems – extract rents to satisfy the wants of elites at the expense of society at large. To target wealth associated with rent extraction, we focus on financial wealth and property wealth *in excess of the value of a household's main residence* – referred to here as investment property wealth. Taxing these types of wealth may contribute to reducing the power of rent-extracting, appropriating systems – or, in Keynes' famous phrase, to the 'euthanasia of the rentier' (Keynes, 1936, chapter 24). Land wealth could also be construed as a source of rent-extraction, but due to data limitations[1], we are not able to include it in our metric.

**Goal number 4: Reduce emissions.** Several authors also argue that wealth taxes can generate environmental benefits, as greenhouse gas emissions and other environmental pressures are highly unevenly distributed along wealth lines (Gore, 2022; Tian *et al*., 2024; Büchs *et al*., 2024b). There are several channels through which wealth taxes could impact emission levels. *Consumption* and *investment* channels are critical. Emissions associated with wealth holdings are substantial, and vary substantially across asset types and countries (Chancel and Rehm, 2025), so wealth taxes could reduce environmental impacts by constraining the emission-intensive investment and consumption activity of the very wealthy. However, due to data constraints, we focus only on the *inequality channel*, i.e. how lower inequality may translate into lower emissions. In this context, Knight *et al*. (2017) find the wealth-inequality elasticity of consumption-based $CO_2$ emissions to be 0.795, meaning that a 1% decrease in wealth inequality is associated with a 0.795% decrease in per capita $CO_2$ emissions.

## 3. Methods and data

### 3.1 Tax models and assessment criteria

While taxes on wealth can exist in many forms, our analysis focuses on taxes on stocks of wealth, rather than on wealth transfers (such as inheritance or gift taxes) or income from wealth (such as taxes on capital gains, profits and interest), in line with most of the wealth-tax literature. This choice is primarily driven by methodological and data constraints. Although the HFCS includes information on inheritance and gifts, these data are widely regarded as unreliable, and no straightforward correction method exists (Alvaredo *et al*., 2017). In addition, taxes on wealth transfers or increases differ in ways that make them less relevant to our goals and criteria. For example, such taxes are levied only when gains are realised, making their incidence irregular and unpredictable, with no direct link between the total stock of wealth and the resulting tax burden or effects. Because we are primarily interested in how a tax would affect outcomes related to stocks of wealth, we focus on taxes directly levied on those stocks.

---

[1] In the HFCS correction procedure outlined in section 3.3, land wealth is split between two variables: land underlying properties is assigned to property wealth, while other land holdings are assigned to non-financial business wealth. As a result, we are unable to accurately model effects on land wealth holdings.

There is also an inherent connection between taxes on stocks of wealth and taxes on income from wealth. A tax on a stock of wealth must ultimately be paid out of income, either from asset sales or from the returns generated by those assets, making a net wealth tax broadly comparable to a tax on capital income. For example, a 1% tax on assets worth €1m with an annual return of 5% is comparable to a 20% tax on the resulting €50,000 in income. However, the two taxes are not equivalent. Revenue from capital income taxes fluctuates with returns, whereas revenue from a tax on the stock of wealth can be more stable if the value of that stock varies less with changes in economic growth. If returns fall to 1% the following year, due to, for example, a financial crisis, the capital income tax rate required to maintain the same revenue would be 100%. Alternatively, if the capital income tax rate remained at 20%, revenue would fall by four-fifths. A tax on stocks of wealth avoids these fluctuations. The tax model with the broadest base and highest burden in our analysis corresponds to a capital income tax of around 70% assuming net wealth of €5m and annual returns of 5%.

We create four tax designs by combining two exemption thresholds with two rate structures, as shown in Table 1 below. We then apply each design to three types of wealth: (i) net wealth, (ii) combined financial and investment property wealth, and (iii) total property wealth, which includes the value of all properties owned, as well as underlying land. This results in twelve combinations. We then use microsimulation analysis to investigate how each of these twelve designs performs against the four goals.

*Table 1: Rates and exemption thresholds applied in the models of four different tax designs*

|  | Exemption threshold | Rate applied to wealth between $90^{th}$ and $95^{th}$ percentile | Rate applied to wealth between $95^{th}$ and $99^{th}$ percentile | Rate applied to wealth above $99^{th}$ percentile |
|---|---|---|---|---|
| Model 1: Lower threshold, lower rates | $90^{th}$ percentile | 1% | 2% | 3% |
| Model 2: Lower threshold, higher rates | $90^{th}$ percentile | 1% | 3% | 5% |
| Model 3: Higher threshold, lower rates | $95^{th}$ percentile | 0% | 2% | 3% |
| Model 4: Higher threshold, higher rates | $95^{th}$ percentile | 0% | 3% | 5% |

By using three tax bases and four tax designs, we can see how the effects of wealth taxes vary when these features differ. Of course, each design has also implications beyond the scope of our analysis. Each of the three tax bases comes with specific advantages and drawbacks. Exempting certain assets (i.e., defining the tax base as anything other than total net wealth) can create incentives for households to shift their holdings toward exempt assets (OECD, 2018; Advani *et al.*, 2020a). For example, if a tax were levied on property wealth, wealthy households might shift their holdings toward other assets that are no less economically or ecologically harmful, or exploit the imprecision of wealth categorisation by, for example, holding property via a financial vehicle so that it is classified as financial wealth instead. Similarly, a tax on financial wealth could have adverse effects on government bond markets – for example, if government bonds become less attractive due to the tax, their price could fall, driving interest rates up (Jackson, 2024). Every wealth tax design has its own implications, and none will be perfect. This again highlights that wealth taxes should not be seen as a magic bullet, but rather as part of a comprehensive package to ensure the tax system as a whole is fair.

The tax rates applied in our analysis are in line with most other studies. For example, Advani *et al.* (2020c) model rates up to 3%, and Apostel and O'Neill (2022) consider rates up to 5%. More ambitious rates also exist in the literature. Kapeller et al. (2023), for instance, simulate rates of up to 90% for the very wealthiest. The tax designs presented in this paper are intended to be policy-relevant and realistic here and now. However, given that rates of return for the very wealthiest can exceed 10% in some contexts (Fagereng *et al.*, 2018), higher rates may likely be necessary to achieve the desired effects in terms of wealth redistribution and emission reduction. The two rate structures and exemption thresholds used in our study are primarily intended to illustrate how results change in response to different designs.

The results of each microsimulation are assessed using criteria aligned with the four post-growth goals, as summarised in Table 2.

*Table 2: Goals and related assessment criteria applied in the modelling*

| Goal | Criterion 1 | Criterion 2 |
| --- | --- | --- |
| Redistribute wealth | Change in the share of wealth held by top 10% and 1%. | Kakwani index of tax progressivity, which indicates the progressivity of the distribution of the tax burden relative to the distribution of wealth. |
| Eradicate extreme wealth | Change in number of households with wealth in excess of €8.9m | Change in number of households with wealth above the 99$^{th}$ percentile. |
| Curb rent extraction | Percentage change in total financial and investment property wealth. | - |
| Reduce emissions | Expected change in $CO_2$ emissions based on Knight *et al.* (2017) elasticity. | - |

## 3.2 Data sources

Estimating the effects of a wealth tax requires microdata on the distribution of wealth. The Household Finance and Consumption Survey (HFCS), provided by the European Central Bank (ECB), is the primary data source underlying our analysis. The HFCS dataset includes data from 22 European countries, (although we remove Croatia, Estonia, Luxembourg and Poland; see Section 3.3 for details), meaning our final dataset contains 18 countries. We use 2017 HFCS survey data, as the most recent available HFCS wave from 2021 may be less reliable due to disruptions caused by the COVID-19 pandemic. The final dataset is based on 80,010 HFCS observations.

Data are also sourced from national accounts and 'rich lists' to correct for the under-representation of the very-wealthy in the HFCS, as explained in the next subsection.

## 3.3 Correcting for under-representation of the very wealthy

Household wealth surveys are subject to two main limitations. First, households at the top of the wealth distribution are less likely to respond than other sampled households (differential unit non-response) (Osier, 2016). The HFCS generally over-samples wealthier households, but research suggests the raw survey data still underestimates the wealth of those at the very top of the distribution (Kapeller et al., 2023). Second, aggregated net household wealth based on surveys is substantially lower than aggregated net household wealth in macro national statistics, and differential unit non-response does not entirely explain this gap (Chakraborty *et al.*, 2019).

It is highly likely that surveys underestimate total net wealth (Waltl, 2022), and therefore wealth tax revenue potential (Apostel and O'Neill, 2022). The ECB has undertaken an effort to correct for wealth survey limitations. While the results of the ECB's correction approach are publicly available, the underlying corrected dataset is not accessible for researchers. We have thus replicated the ECB's correction procedure, drawing on the ECB's methodological reports (Engel *et al.*, 2022; ECB, 2024). As some input data is not publicly available, we are unable to fully replicate the ECB's work, but our results are in general very close to their estimates.[2]

In what follows, we describe the main steps of the correction procedure and highlight important differences between our replication and the ECB's original work.

**Step 1: Adjust HFCS weights**. The HFCS target population excludes households in collective institutions (e.g., prisons). To correct for this, weights are adjusted so that the adjusted HFCS weights sum up to the total number of households in each country.

**Step 2: Link HFCS variables to their National Accounts counterparts**. While most HFCS variables can be easily linked to their National Accounts counterparts (ECB, 2020), there are some exceptions.[3] Here we consistently follow the ECB's approach to dealing with these exceptions. In some cases, the ECB relies on non-public data to adjust National Accounts aggregates before the linking procedure. Since these adjusted aggregates are publicly available, we adopt them in our replication procedure.

**Step 3: Correct for underreported deposits**. The ECB corrects deposits that are implausible considering a household's total income or assets.[4] Since the ECB (2024) methodological note is vague on the exact procedure, we adopt the approach described in Engel *et al.* (2022).

**Step 4: Estimate the top tail and sample rich households**. It is well-known that household wealth is Pareto distributed at the top. The country-specific shape of this Pareto distribution is estimated on a dataset which combines the richest households from the HFCS (in the ECB baseline, those households with wealth above €1m) and very-wealthy households from 'rich lists' such as Forbes. As the richest households in the HFCS are typically much poorer than the poorest households in a rich list, the gap between the HFCS and the rich list is filled by sampling households from the estimated Pareto distribution. Here we diverge from the ECB given that the country-specific rich lists used by the ECB are not publicly available. Instead, we rely on the Forbes list and, where there are no Forbes rich list observations close to the survey reference period, the European Rich List Database (ERLB) (Disslbacher

---

[2] Both the top 5% and top 10% net wealth shares are on average only 0.4%-points larger in our estimates than in the DWA. For the Netherlands and Greece our estimates are substantially larger than the ECB's results (by 10%-points and 7%-points for the top 5% net wealth shares, respectively). In the case of the Netherlands, this divergence can be explained by the fact that the Dutch DWA data are based on administrative data rather than HFCS microdata. The reason for our larger estimate for Greece is not clear. When excluding the Netherlands and Greece, our wealth share estimates are on average 0.6%-points smaller than the ECB for the top 5% and 0.4%-points smaller for the top 10% net wealth shares.

[3] For example, business wealth (an HFCS variable) can be mapped to either unlisted shares and other equity (if the business is regarded as a corporation or a quasi-corporation) or not be recorded separately (i.e. no distinction is made between business and household assets, if the household is considered to be a producer household).

[4] There is also the issue that business deposits are classified as business wealth in the HFCS but as deposits in the National Accounts. We also follow the ECB's correction for this issue.

*et al.*, 2022).[5] For four countries (Croatia, Estonia, Luxembourg and Poland), either the rich list data or the ECB DWA do not exist. These countries are not included in our final sample. Our final sample contains 80,010 original HFCS observations, plus between 81,422 and 87,697 imputed observations. The number of imputed observations varies by implicate due to differences in the pareto distributions.

**Step 5: Specify the portfolio allocation at the top**. While the HFCS contains detailed information on the portfolio allocation of households, rich lists typically only report on total net household wealth. Moreover, for the sampled top households, only total net wealth is available. The ECB first estimates country-specific liability to net wealth ratios. These liabilities estimates can then be used to infer gross total wealth. Gross total wealth is then allocated to different asset classes using external information on top wealth portfolios.[6] Unfortunately, the ECB (2024) methodological note is vague on the liability correction procedure, so we take the average of the (relatively narrow) cross-country range they mention.

**Step 6: Proportional rescaling**. To fully align the corrected HFCS data with the National Accounts, HFCS asset categories are rescaled to their linked (in Step 2) National Accounts counterparts. For liabilities, the rescaling procedure is adapted to ensure that net negative households do not end up with more negative net wealth than they had before (as the standard procedure leads to implausible outliers).

The ECB does not implement Step 3 and/or Step 4 for a small number of countries, as in these cases such corrections were not deemed necessary, either because the deposits are considered not to be underreported, or the HFCS survey is considered to already capture the top of the wealth distribution very well. We adopt the ECB's approach for these countries.

Table 3 shows the shares of wealth held by the top 10%, 5% and 1% for our sample of 18 countries in the original, uncorrected HFCS data, and the corrected HFCS data.

*Table 3: Top wealth shares using uncorrected and corrected data*

| Share of wealth held by the top… | Uncorrected HFCS data | Corrected HFCS data |
|---|---|---|
| 10% | 52.1% | 57.4% |
| 5% | 38.2% | 44.6% |
| 1% | 18.2% | 25.8% |

*Source: Own calculations based on HFCS (2017).*

### 3.4 Microsimulation methods

After correcting the HFCS data to more accurately represent the true wealth distribution, we performed static microsimulation analysis to simulate the effects of the different tax designs. Our analysis is conducted at the household level.

---

[5] Moreover, we consistently sample to fill the gap between HFCS and rich list observations, whereas the ECB does not apply sampling in some cases (as they likely have more rich list observations) or additionally samples at HFCS or rich list wealth levels (the precise criteria are unclear, so we decided not to implement this additional sampling).

[6] ECB (2024) does not report a precise breakdown of the portfolio allocation, so we rely on the breakdown provided in Engel et al. (2022). We do implement the country-specific portfolio allocation adjustments (relative to the baseline) detailed in the ECB (2024) report, to the extent that such adjustments are possible based on publicly available information

We recognise that these 18 countries do not represent a political community capable of jointly deciding wealth-tax policy. We do not suggest that this sample should constitute the geographical scope of wealth taxation. Rather, we aim to provide, as closely as possible within data limitations, an illustration of the potential of wealth taxation in Europe. We analyse the dataset as a whole, rather than the individual countries, to ensure that exemption and marginal-rate thresholds are consistent across the sample. As a result, the simulated taxes are redistributive within the whole sample, rather than within each country.

Revenue estimates are generated by summing taxable wealth – i.e., the wealth subject to taxation under each design – and multiplying by the applicable tax rate(s). Where taxes are levied on specific types of wealth, the percentile thresholds refer to that specific type of wealth, *not* net wealth.

Analysis for **goal 1** (redistribute wealth) involved calculating the change in the share of total wealth held by the top 10% and the top 1%, before and after the tax. Additionally, we compute the Kakwani index of tax progressivity (Kakwani, 1977), which measures the difference between the concentration index of tax payments (i.e., the distribution of the tax burden across the wealth distribution) and the Gini coefficient (in this case, for net wealth). A positive Kakwani index indicates that the tax payments are more concentrated at the top than wealth itself, implying a progressive tax. The higher the index, the more progressive the tax.

For **goal 2** (eradicate extreme wealth), we count the number of households with net wealth exceeding €8.9m (the most popular absolute threshold from Balata *et al*. (2025)) and the number with net wealth greater than the 99$^{th}$ percentile (approximately €2.4m – a relative threshold roughly in line with Dutch public perceptions of extreme wealth from Robeyns *et al.* (2021)), before and after the tax, and calculate the change.

For **goal 3** (curb rent-seeking), we measure the percentage change in the total amount of financial and investment property wealth. The procedures for doing so differ depending on the tax base. When the tax base is net wealth, we assume that each household's wealth portfolio composition remains constant. Therefore, we subtract from each household's net wealth an amount of financial and investment property wealth equal to their total tax liability multiplied by the share of their wealth held in these asset types. Because wealthier households tend to hold a larger proportion of their wealth in financial assets and investment property – and are more likely to exceed the exemption thresholds – taxing net wealth still reduces the amount of wealth held in these asset types. When the tax base is total property wealth, we have to consider the impact of the procedure that corrects for the under-representation of the very wealthy. As described in Section 3.3, the data-correction procedure only adjusts aggregated property wealth. To separate out investment property, we used the raw HFCS data to determine the proportion of total property wealth held in investment property for each wealth decile and for the top percentile. We then apply these proportions to the corrected aggregate property wealth variable to reconstruct separate asset categories. This approach has limitations, since the asset composition of very-wealthy households is sometimes adjusted to align with national accounts data. However, it allows us to estimate the reduction in investment property wealth under a property wealth tax by multiplying the total tax revenue by the share of total property wealth held in investment property. When the tax base is financial and investment property wealth, the calculation is more direct: we simply deduct total tax revenue from total pre-tax financial and investment property wealth.

For **goal 4** (reduce $CO_2$ emissions), we use the wealth-inequality elasticity of consumption-based $CO_2$ emissions from Knight *et al*. (2017), which the authors estimate to be 0.795. In their analysis, wealth inequality is measured by the share of wealth held by the top 10%. To estimate the expected direct reduction in $CO_2$ emissions resulting from a decrease in wealth inequality, we multiply the change in the top 10%'s wealth share by 0.795.

## 4. Results

### 4.1 Wealth distribution in 18 European countries

To begin, we explore how different types of wealth are distributed across households in our sample of 18 European countries. Table 3 shows the amount of each wealth type (in €) held at selected percentiles, and Figure 1 presents the corresponding relative generalised Lorenz curves for the three types of wealth.[7]

*Table 4: Wealth holdings by percentile and wealth type (in €)*

| Wealth type | 50$^{th}$ percentile | 75$^{th}$ percentile | 90$^{th}$ percentile | 95$^{th}$ percentile | 99$^{th}$ percentile | Gini coefficient |
|---|---|---|---|---|---|---|
| Net wealth | 112,620 | 308,413 | 629,352 | 973,265 | 2,406,940 | 0.70 |
| Financial and investment property wealth | 42,181 | 141,487 | 407,042 | 673,575 | 1,710,737 | 0.79 |
| Total property wealth | 102,373 | 255,552 | 453,947 | 652,011 | 1,333,699 | 0.67 |

*Source: Own calculations based on 18 countries of the HFCS (2017). Sample consists of 80,010 HFCS observations. Reported results are averages from the five implicate datasets.*

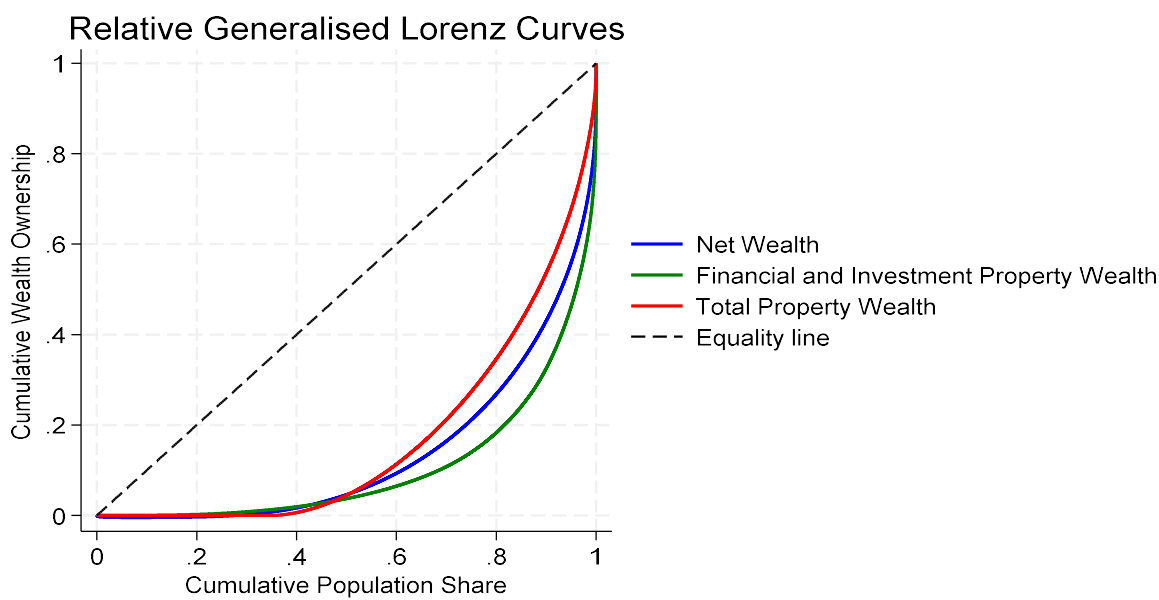

*Figure 1: Relative generalised Lorenz curves for the three different types of wealth. Source: Own calculations based on 18 countries from the HFCS (2017). Sample consists of 80,010 HFCS observations. Reported results are averages from the five implicate datasets.*

---

[7] Based on the 1$^{st}$ implicate dataset. Results for other implicate versions are very similar.

This analysis shows that the inequality of wealth distribution differs by type of wealth. Financial and investment property wealth is the most unequally distributed. Net wealth is somewhat less unequal. Total property wealth, though still highly unequal, is the most evenly distributed.

## 4.2 Revenue

This section presents estimates of how much revenue (€bn) could be generated following the first implementation of each tax design (Figure 2). The largest amount would be raised by a tax on net wealth with a lower threshold and higher rates – €600bn, or approximately 13% of the total tax revenue for our sample of countries in 2017 (OECD, no date). A tax on total property wealth with a higher threshold and lower rates would yield the lowest revenue, at €111bn. It is important to note, however, that these estimates do not account for avoidance and evasion. They also represent only the expected revenues from the first time the tax is levied. These limitations are discussed in Section 5.

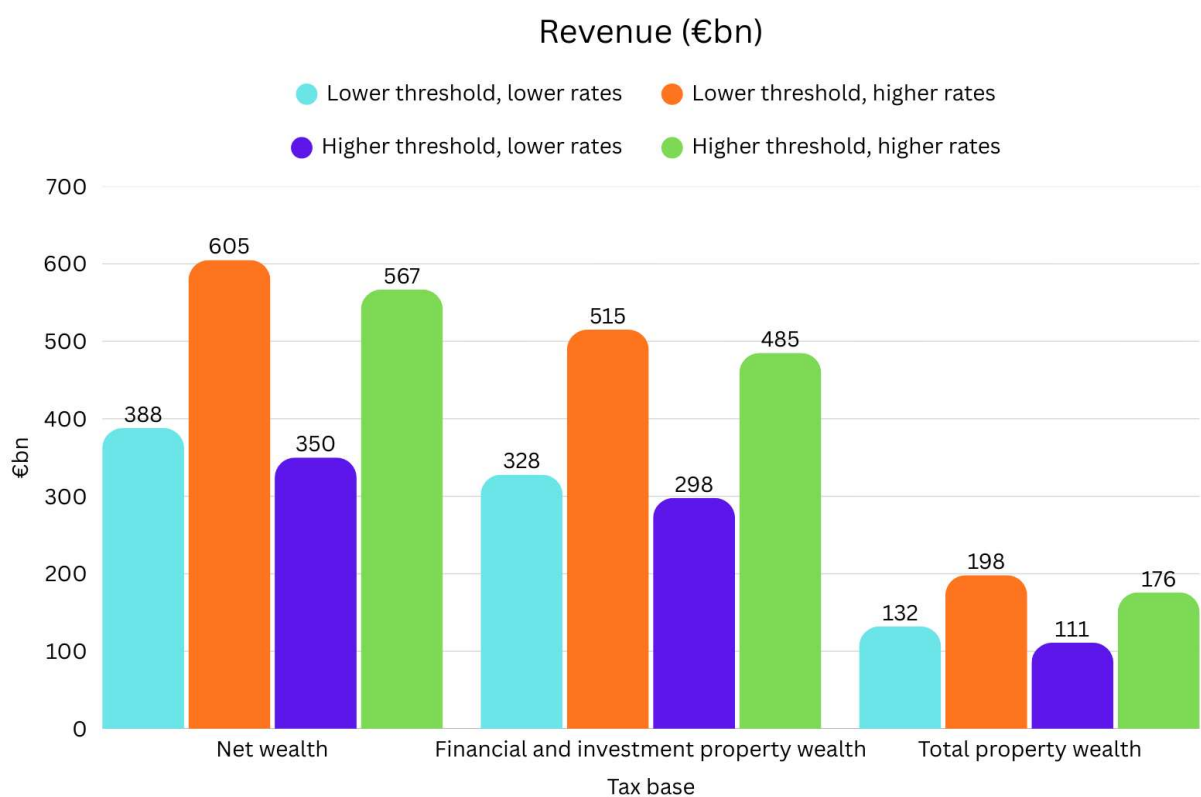

*Figure 2: Revenue estimates (€bn) for each tax design. Source: Own calculations based on 18 countries from the HFCS (2017). Sample consists of 80,010 HFCS observations. Reported results are averages from the five implicate datasets.*

## 4.3. Effects of different tax designs in relation to post-growth goals

### 4.3.1 Goal 1 – Wealth redistribution

This section summarises our results regarding the wealth-redistributing potential of the different tax designs. Figure 3 shows the percentage-point reduction in the share of wealth held by the top 10%, and Figure 4 the percentage-point reduction in the share held by the top 1%, after the implementation of the taxes. The most effective tax for reducing the top 10% share of wealth is a net wealth tax with higher rates and a lower threshold. This design would be expected to reduce the top 10% wealth share by 0.56 percentage points in the first year. To reduce the top 1% share, the most effective design is also a net wealth tax with higher rates, but this time with a higher exemption threshold. This would be expected to reduce the wealth share of the top 1% by 0.73 percentage points in the first year. The least effective tax design to reduce wealth inequality, by these measures, is one levied on total property wealth, especially with low tax rates.

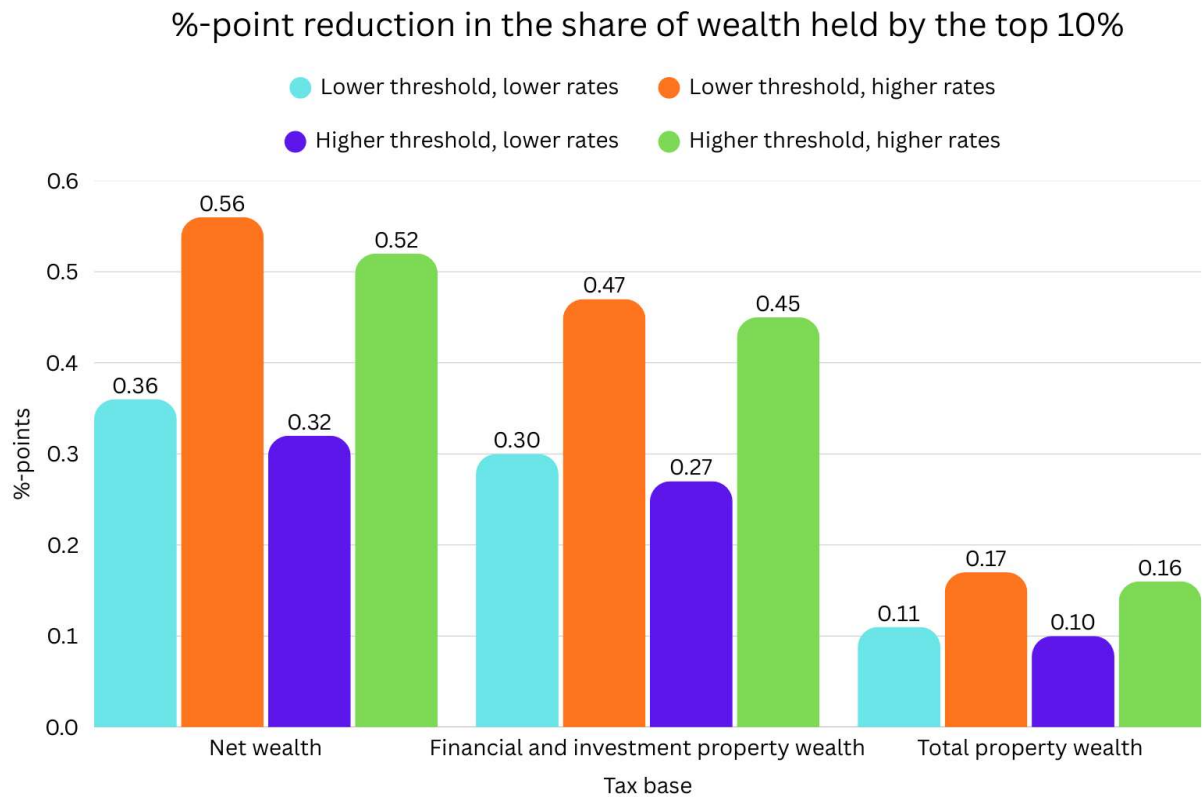

*Figure 3: Percentage point reduction in the share of wealth held by the top 10%. Source: Own calculations based on 18 countries from the HFCS (2017). Sample consists of 80,010 HFCS observations. Reported results are averages from the five implicate datasets.*

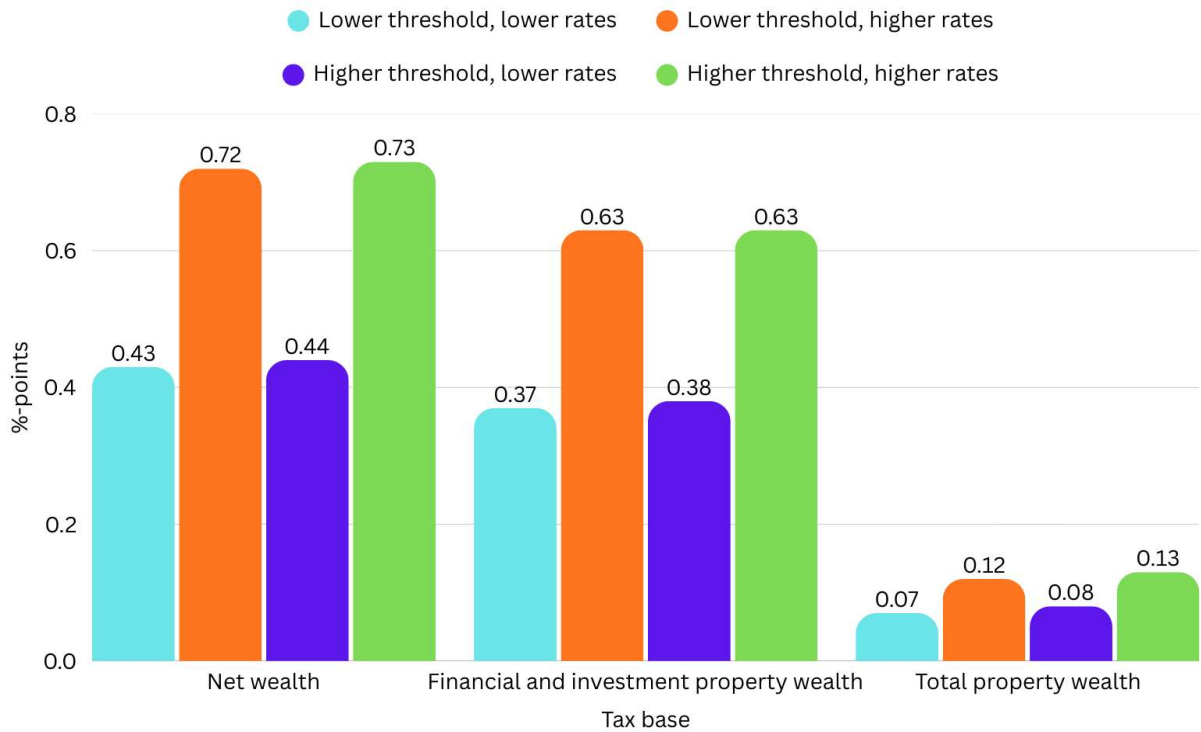

*Figure 4: Percentage point reduction in the share of wealth held by the top 1%. Source: Own calculations based on 18 countries from the HFCS (2017). Sample consists of 80,010 HFCS observations. Reported results are averages from the five implicate datasets.*

Figure 5 shows the Kakwani index for each tax design. The Kakwani index compares the distribution of tax payments with the distribution of wealth. Based on this measure, the most progressive tax is one with a higher exemption threshold levied on net wealth, or taxes with higher exemption thresholds and higher rates on financial and investment property wealth. The least progressive taxes are those levied on total property wealth, especially when the exemption threshold is low. The Kakwani index shows taxes with higher thresholds to be more progressive, because the tax payments are concentrated higher up the wealth distribution, however, as shown in figure 3, these aren't always the most effective at reducing top wealth shares.

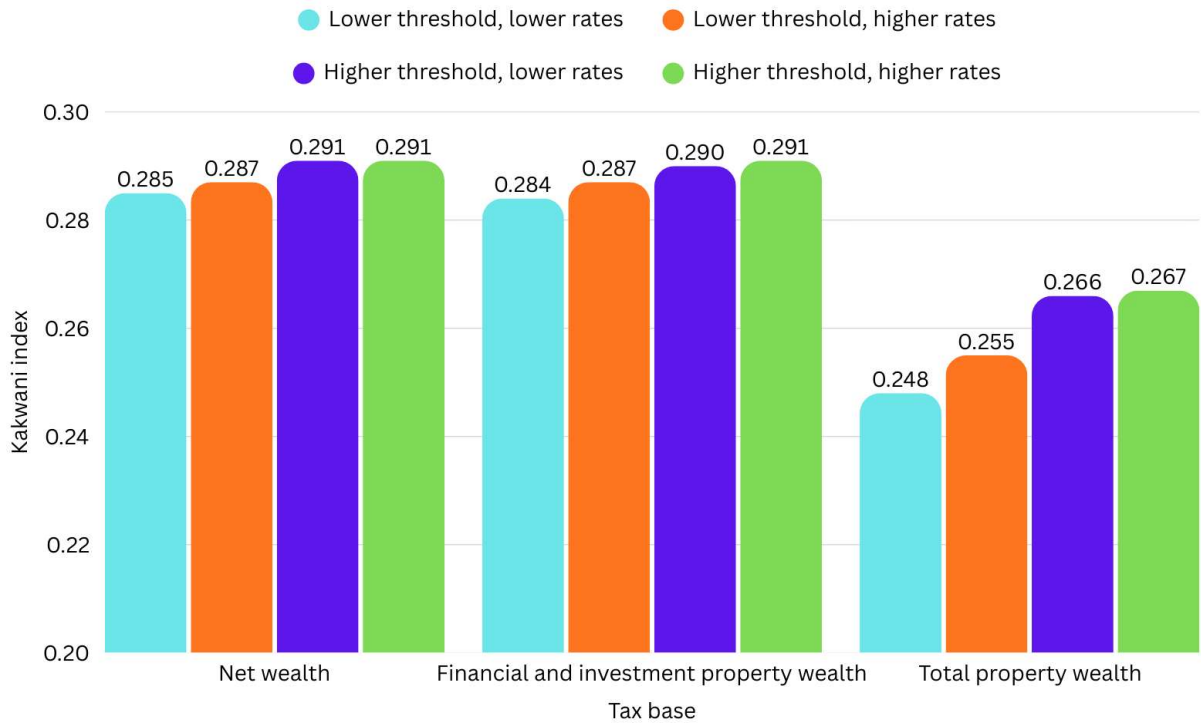

*Figure 5: Kakwani index for each tax design. Source: Own calculations based on 18 countries from the HFCS (2017). Sample consists of 80,010 HFCS observations. Reported results are averages from the five implicate datasets.*

### 4.3.2. Goal 2 – Eradicate extreme wealth

This section presents our results on the reduction of extreme wealth, using two different thresholds. Figure 6 shows the change in number of households with net wealth exceeding €8.9m – the most popular absolute threshold for 'extreme' wealth, according to Balata *et al.* (2025) – after the implementation of the tax. The most effective tax design is a net wealth tax with higher rates and a lower exemption threshold, which reduces the number of households above this level by around 11,000 (just over 7%) in the first year. At the other end of the scale, a total property wealth tax with lower rates and a higher threshold results only in a reduction of 670 households, or about 0.4%.

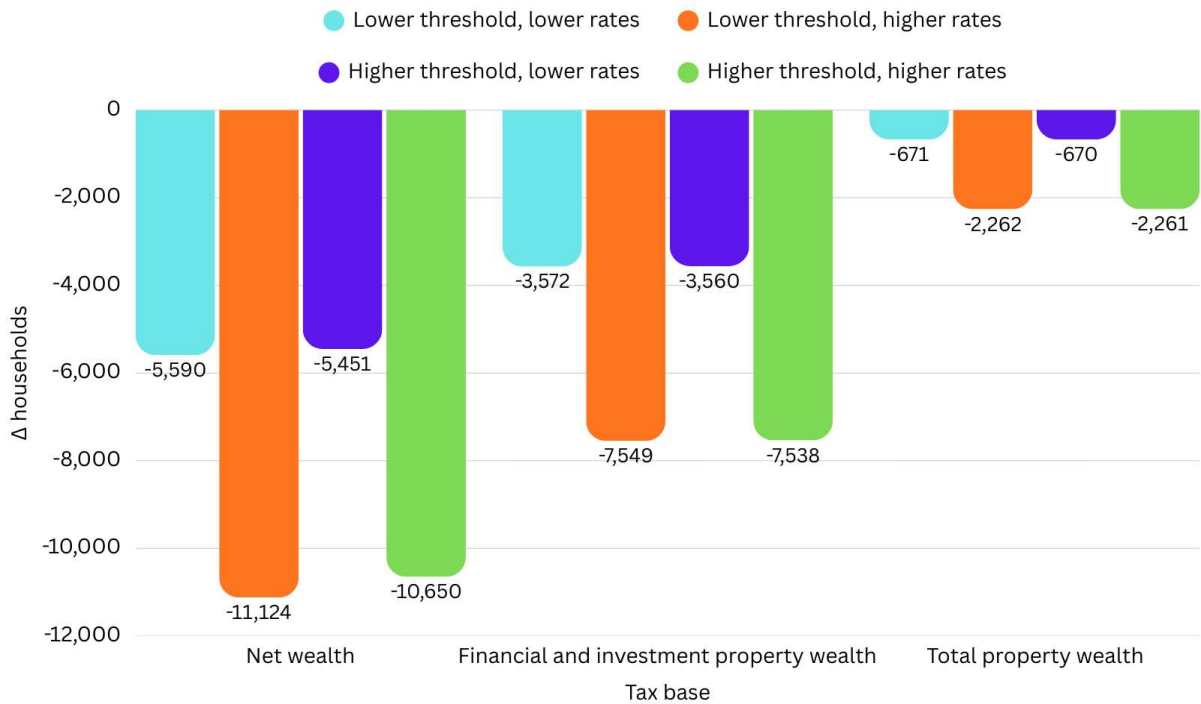

*Figure 6: Change in number of households with net wealth exceeding €8.9m. Source: Own calculations based on 18 countries from the HFCS (2017). Sample consists of 80,010 HFCS observations. Reported results are averages from the five implicate datasets.*

Figure 7 shows the change in the number of households with net wealth exceeding €2.4m (the 99th percentile) after implementation. Using this relative threshold, the same tax design remains the most effective: a net wealth tax with higher rates and a lower exemption threshold reduces the number of households above the 99th percentile by around 77,000 (just under 5%) in the first year. In contrast, a total property wealth tax with lower rates and a higher threshold reduces this group by around 26,000 (or 1.7%).

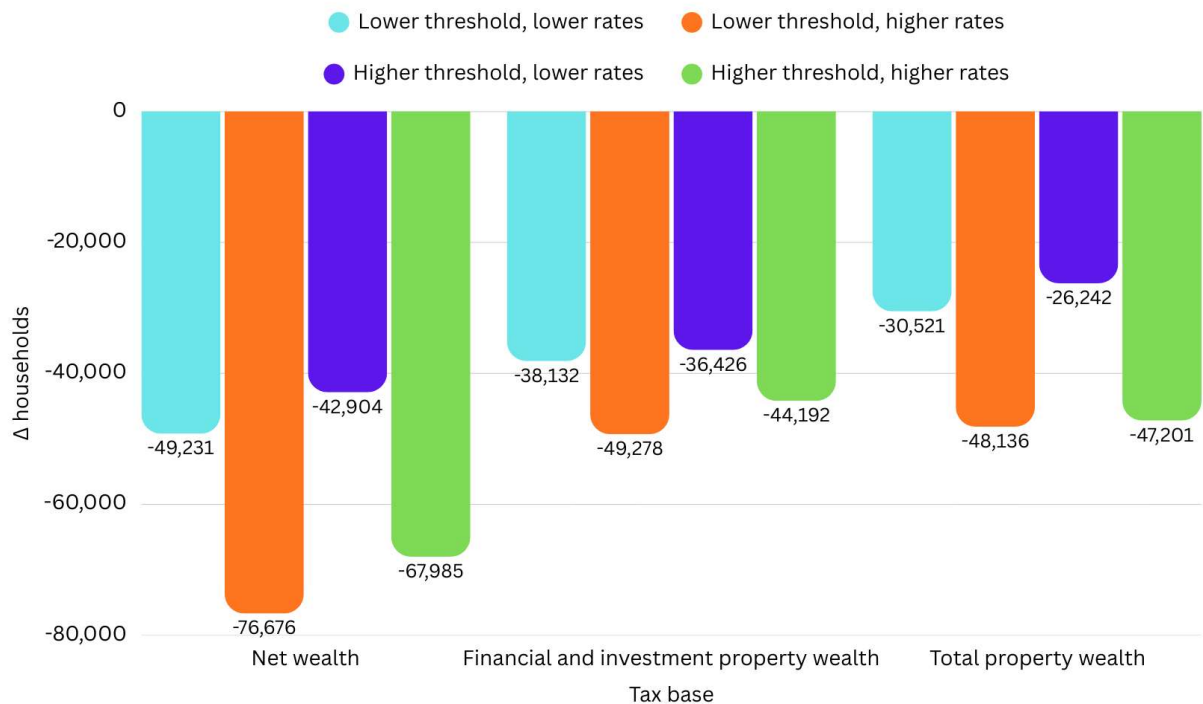

*Figure 7: Change in number of households with net wealth in excess of the 99th percentile. Source: Own calculations based on 18 countries from the HFCS (2017). Sample consists of 80,010 HFCS observations. Reported results are averages from the five implicate datasets.*

### 4.3.3 Goal 3 – Curb rent extraction

This section summarises our results on how the different tax designs may affect the composition of wealth holdings in the economy. Figure 8 shows the percentage change in the amount of wealth held in financial assets and investment property following implementation of each tax design. If the goal is to shift the economy away from rent-seeking activities based on finance and investment property, then a tax directly targeting these types of wealth is, unsurprisingly, the most effective option. A higher-rate, lower-threshold tax on financial and investment property wealth reduces holdings of financial and investment property wealth by 1.72% (or approximately €500bn). This reflects the direct effect of wealth being relinquished through taxation. Additional indirect effects are also likely. For example, such a tax may incentivise shifts in wealth portfolios toward tax-exempt assets (i.e., productive assets in the 'real' economy), further reducing the share held in financial and investment property wealth.

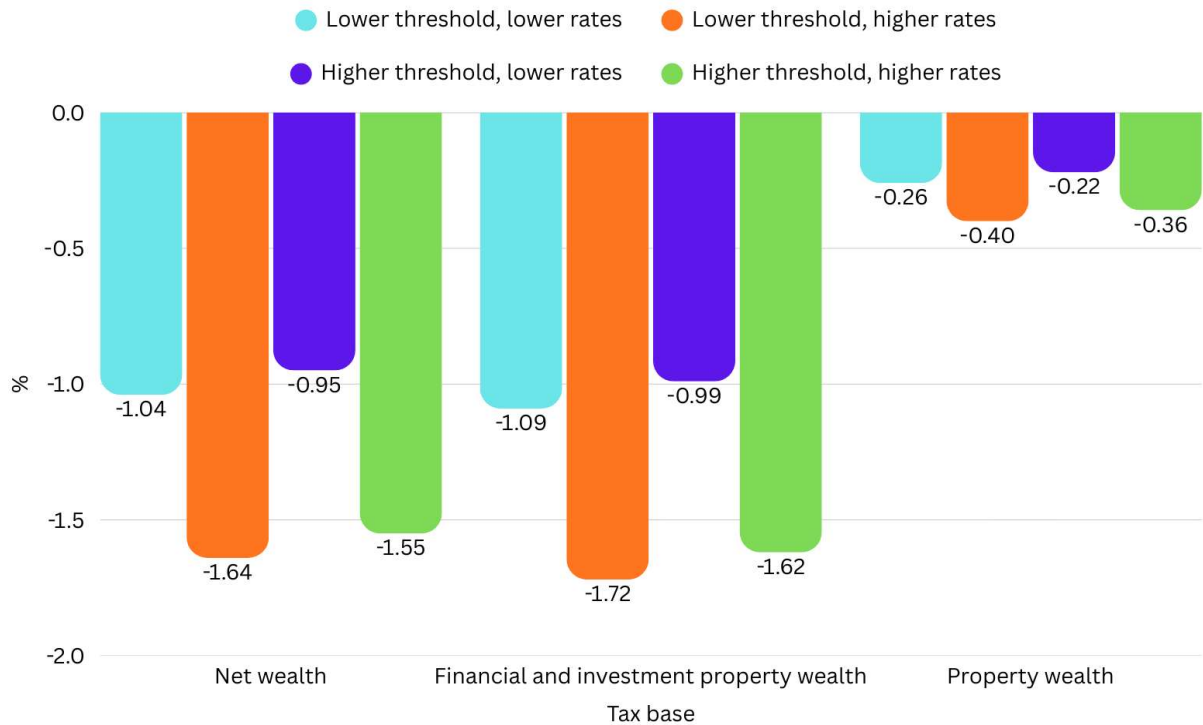

*Figure 8: Percentage change in amount of wealth held in financial assets and investment property after the implementation of each tax design. Source: Own calculations based on 18 countries from the HFCS (2017). Sample consists of 80,010 HFCS observations. Reported results are averages from the five implicate datasets.*

### 4.3.4 Goal 4 – Reduce $CO_2$ emissions

This section summarises the potential $CO_2$ emission reductions associated with each tax design. Figure 9 shows the expected percentage change in $CO_2$ emissions for each tax design, based on the change in the wealth share of the top 10% and the elasticity calculated by Knight *et al*. (2017). According to this analysis, a tax on net wealth with higher rates and a lower exemption threshold is the most effective, reducing $CO_2$ emissions by 0.77% in the first year. A tax on total property wealth with lower rates and a higher threshold is the least effective, reducing $CO_2$ emissions by 0.14%. However, all of these estimates rely on a single associative elasticity from Knight *et al*. (2017) and should therefore be interpreted with caution.

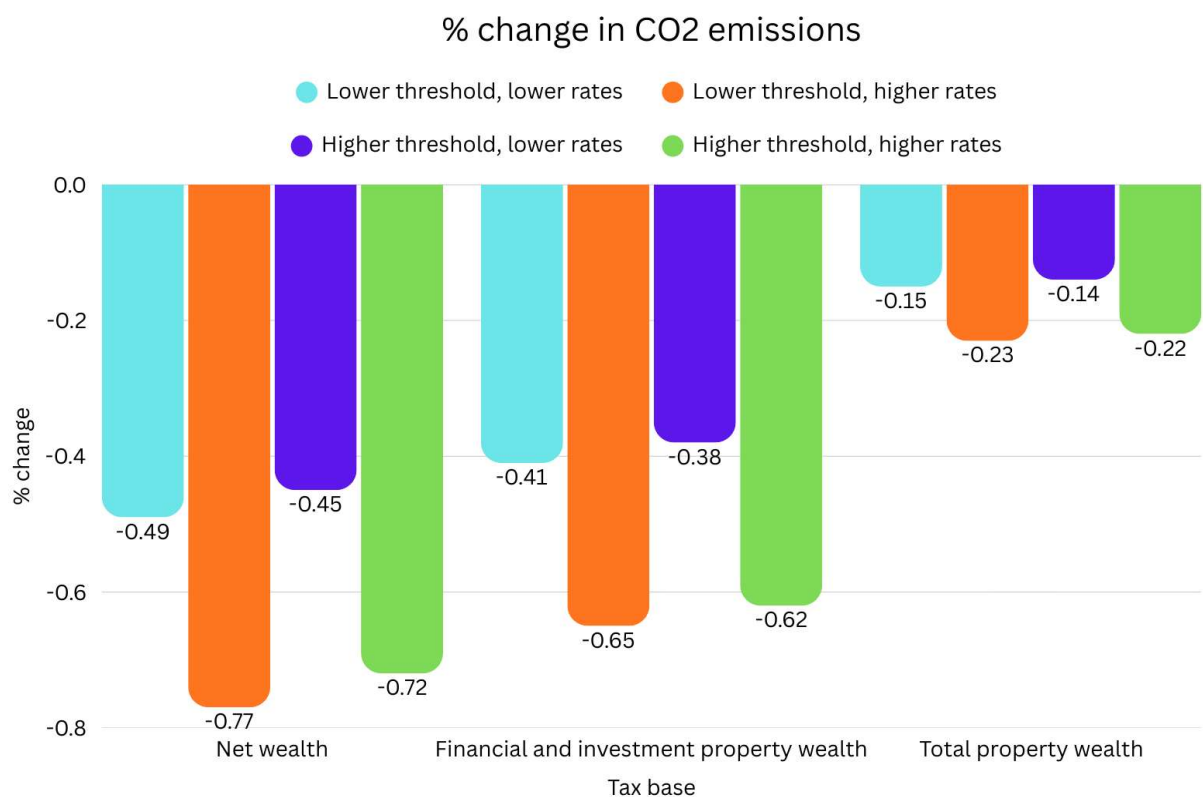

*Figure 9: Expected % change in CO2 emissions for each tax design. Source: Own calculations based on 18 countries from the HFCS (2017). Sample consists of 80,010 HFCS observations. Reported results are averages from the five implicate datasets.*

# 5. Discussion

In this section, we place our findings in the context of the broader literature and discuss their implications in more detail. To put our revenue estimates into perspective, government spending on social protection in our country sample totalled around €3 trillion in 2017 (OECD, no date). A wealth tax with higher rates and lower thresholds could raise up to €605bn – around 20% of this expenditure (with actual revenues likely being somewhat lower due to evasion/avoidance) – or, from an MMT perspective, create substantial macroeconomic space for increased spending on social and ecological objectives. Therefore, our analysis shows that wealth taxes can play an important role in enabling post-growth welfare states.

Our analysis regarding wealth redistribution and emission reduction is roughly in line with the limited number of existing studies. Our estimate that the share of wealth held by the top 1% could be reduced by 0.73% (under a net wealth tax with higher rates and a higher threshold) is within the range of 0.16% to 0.84% given by Apostel and O'Neill (2022). Our estimate that $CO_2$ emissions could be reduced by up to 0.77% is also similar to the estimate of 0.1%-0.6% from Apostel and O'Neill (2022).

## 5.1 Evaluating tax designs against post-growth goals

Our analysis shows that the most effective tax design varies depending on the goal of the tax. If the goal is to maximise revenue or the amount of wealth recirculated, reduce the share of wealth held by the top 10%, reduce the number of households with 'extreme' wealth, or reduce $CO_2$ emissions, then a tax on net wealth with a lower threshold and higher rates is most effective. If the goal is to maximise the progressivity of the tax or reduce the share of wealth held by the top 1%, a tax on net wealth with a higher threshold and higher rates yields the best results. If the goal is to reduce rent extraction, a tax on financial and investment property wealth with higher rates and a lower threshold performs best.

## 5.2 Trade-offs

Our analysis also highlights conflicts between the goals. Figure 10 presents these visually showing the relative effectiveness of taxes on different kinds of wealth in relation to the four goals, as well as total revenue. Figure 10 corresponds to the 'higher rates, lower threshold' models. The pattern is similar for other models. Outcomes for each of the criteria are indexed against the most effective option, which is indexed to 100. Where there are multiple criteria per goal, an average is taken.

As figure 10 shows, a tax on net wealth would be the most redistributive. A tax on financial and investment property wealth would be most effective at shifting the economy away from rent-seeking activities but would not be as effective at redistributing overall wealth. However, if rent extraction was reduced, redistributive taxation would have less work to do. Evidence from the UK also indicates higher levels of public support for taxes on financial and investment property wealth than other types of wealth (Rowlingson *et al.*, 2020), based on perceptions that this type of wealth is more likely to be 'unearned'. However, while a tax on financial and investment property wealth must form part of a comprehensive policy package to curb rent extraction, it is unlikely to be an adequate response on its own in an otherwise unchanged system. As (Stratford, 2020, p.1) points out, rent extraction – 'an economic reward which is sustained through control of assets that cannot be quickly and widely replicated, and which exceeds proportionate compensation for the labour of the recipient' – is a structural symptom of an economy built on often privatised control over scarce assets. Varoufakis' (2024) account of technofeudalism reinforces this: because today's dominant rents increasingly stem from the structural power of digital platforms and cloud capital, taxation alone is insufficient, meaning that even a well-designed wealth tax must be complemented by broader reforms, such as regulating or breaking up dominant platforms, placing key digital infrastructures under public or commons ownership, and limiting the monopolisation of data and algorithms.

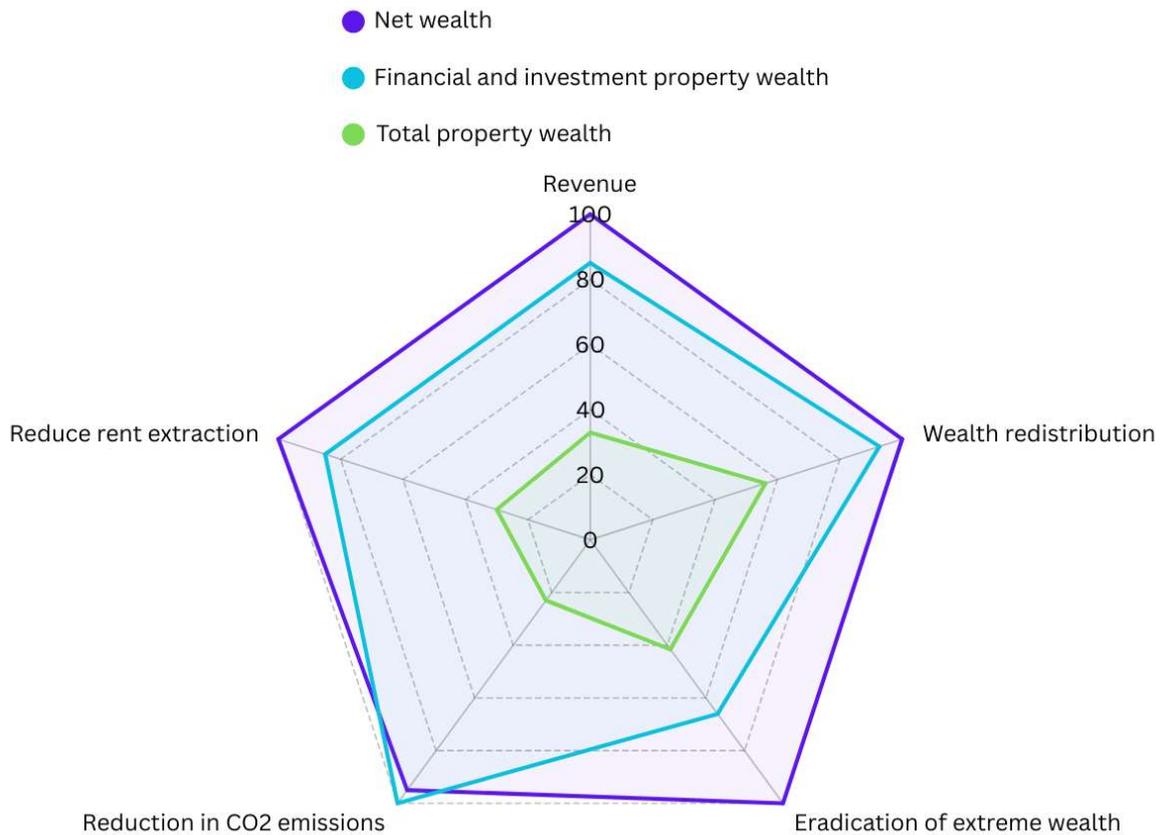

*Figure 10: Radar chart showing relative effectiveness of taxes on different kinds of wealth. Based on LT, HR models.*

Trade-offs between goals and tax designs emphasise the need for holistic policy portfolios (Bärnthaler *et al.*, 2026) – there is no single policy that will achieve all the goals of a post-growth transition. For instance, a wealth tax could be levied on net wealth to maximise redistribution, whilst rent-extraction could be addressed through rent controls, de-commodified provisioning, and worker-owned companies. The latter could be encouraged by allowing wealth tax liabilities to be paid in the form of transferring shares to workers (Parrique, 2019, p.692). At the same time, wealth taxes can help set the ceiling of equitable production and consumption corridors and create the macroeconomic space needed for policies like Universal Basic Services, which establish the floor of universal need satisfaction (Bärnthaler and Gough, 2023; Büchs, 2021b). Sequencing, however, will also be important: (Stratford, 2020, p.3) argues that if environmental protections that limit opportunities for expanding production precede action on rent extraction, they may exacerbate rent-seeking behaviour and increase economic inequality, 'as those seeking to increase their income and wealth may be tempted […] to make more powerful claims over the spoils from *existing* production'.

Despite our analysis being static, effectively conceptualised as a one-off tax (see limitations), trade-offs relating to the dynamic effects of taxation also need to be considered. For instance, if a wealth tax were levied at a rate sufficiently high to prevent the accumulation of wealth, revenue would fall in subsequent rounds of taxation. However, if rates were set below average rates of return, this would not be the case. This highlights how trade-offs between goals may become more pronounced over time. If, for ecological or social reasons, high levels of private wealth are viewed as something to be abolished via taxation, there will be greater effects in terms of wealth redistribution and emission reduction, but the tax will eventually undermine its own revenue base. Conversely, if wealth is treated primarily as a useful source of tax revenue and allowed to continue growing, revenue will be more stable, but at the cost of weaker distributional and environmental outcomes.

Finally, there are also trade-offs to consider regarding liquidity. A tax with an exemption threshold at the 90$^{th}$ percentile will generate more revenue and redistribute more wealth than one with an exemption threshold at the 95$^{th}$ (or higher) percentile. However, it will also result in more households facing liquidity concerns (Loutzenhiser and Mann, 2020). This also depends on the type of wealth taxed – financial wealth is generally more liquid (Advani *et al.*, 2020b), so if the tax were levied on financial and investment property wealth, liquidity concerns would be mitigated, albeit at the expense of revenue. Careful planning is therefore required to ensure that liquidity issues are not excessive, but if the goal is a more equal distribution of assets, it may not be undesirable for the wealthiest households to sell some of their holdings to pay the tax bill.

## 5.3 Limitations

The analysis in this article has several limitations. Firstly, the analysis is static. This means the results only represent the immediate, or 'day after', effects of a wealth tax. Our analysis is thus, in effect, conducted as if it were a one-off wealth tax, whereas ideally research should examine the impacts of recurring taxes.

Secondly, we do not consider tax avoidance or evasion in our empirical analysis because existing estimates of the elasticity of taxable wealth in response to changes in wealth tax rates vary enormously – by a factor of 800 (Advani and Tarrant, 2021) – reflecting methodological differences and the strong influence of policy, cultural, and regulatory contexts (OECD, 2018; Advani and Tarrant, 2021). Meaningful estimates for our tax designs are therefore not feasible, as these designs would be unprecedented in geographical scope, rate and threshold structure, and tax base. Nonetheless, design choices can limit avoidance and evasion – for example, taxing on the basis of citizenship rather than residency, or taxing worldwide assets to reduce capital mobility (OECD, 2018; Advani and Tarrant, 2021). Improvements in information technology and international data-sharing (Saez and Zucman, 2022) have already reduced evasion threefold over the past decade (Alstadsaeter *et al.*, 2024), and wealth tax returns could be pre-filled by the tax authorities to further narrow avoidance opportunities (Saez and Zucman, 2019). It is also worth noting that the social and political-economic context implied by a post-growth transition is very different from the current political-economic arrangement. Behavioural responses to taxation, and the ability of authorities to deal with such responses, may therefore differ.

Thirdly, there are myriad interaction effects that cannot be fully anticipated. For example, a tax on wealth stocks may have repercussions for the revenue from other taxes, as such a tax would partially replace, rather than simply add to, the revenue from capital income or inheritance taxes (because if stocks of wealth have been taxed, income from wealth, and the amount of wealth transferred upon death, will be comparatively lower). Some potential interaction effects could also be positive. For example, Büchs *et al*. (2021) explore different ways of 'recycling' revenues from carbon taxes in European countries. They show that by using revenues to equitably expand the provisioning of green public services such as renewable energy and public transport, greater environmental and social benefits can be realised. The same could be achieved with wealth tax revenues.

Finally, although their potential to generate more growth-independent tax revenue is one potential justification for the inclusion of wealth taxes in post-growth policy portfolios, we are not able to assess this goal based on the dataset and methods we are using here. This is an important opportunity for future research.

# 6. Conclusion

Wealth taxes are a commonly proposed policy within the post-growth literature, yet an assessment of their performance against post-growth goals and empirical estimates of their potential effects remain limited. We contribute by using microsimulation to explore, for the first time, how twelve concrete wealth-tax configurations perform against four key goals of a post-growth transition: reducing inequality, eradicating extreme wealth, curbing rent extraction, and reducing $CO_2$ emissions.

Our results show that no single tax design advances all goals equally. For example, taxes on net wealth are the most effective for redistribution, extreme wealth reduction, and $CO_2$ emission decreases, while taxes on financial and investment property wealth are the most effective at reducing rent-seeking activity. These trade-offs highlight that wealth taxes cannot serve as a standalone solution; instead, they must be embedded within broader eco-social policy packages that complement their strengths and mitigate their limitations.

In an era of polycrisis, where ecological breakdown and extreme inequality are mutually reinforcing, there is an urgent need for policies that transform the underlying political-economic dynamics. Our analysis demonstrates that well-designed wealth taxes can make a meaningful contribution to this transformation by recirculating wealth, constraining harmful forms of accumulation, and supporting more equitable and sustainable provisioning systems. Their transformative potential, however, depends on being coupled with wider reforms – such as regulatory measures to curb rent extraction, democratise ownership, and strengthen universal eco-social provisioning – that together can support a just and sustainable post-growth transition.

# CRediT authorship contribution statement

**Thomas Webb:** Conceptualisation, Methodology, Formal Analysis, Investigation, Writing – Original Draft, Writing – Review and Editing, Visualisation.

**Arthur Apostel:** Methodology, Formal Analysis, Writing – Original Draft, Writing – Review and Editing.

**Milena Büchs:** Conceptualisation, Methodology, Writing – Review and Editing, Supervision, Funding Acquisition.

**Richard Bärnthaler:** Conceptualisation, Writing - Review & Editing, Supervision.

# Funding acknowledgement

Thomas Webb and Milena Büchs have received funding through the Horizon Europe project MAPS (Models, Assessments and Policies for Sustainability), grant agreement number 101137914.

Arthur Apostel is the recipient of a PhD Fellowship grant from the Research Foundation Flanders (11PP924N).

## Acknowledgements

We would like to thank members of the Economics and Policies for Sustainability (Econopol) research group at the University of Leeds, and the MAPS consortium for their helpful comments during the production of this paper, as well as two anonymous reviewers for their valuable comments.

## References

Advani, A., Bangham, G. and Leslie, J. (2020b) *The UK's wealth distribution and characteristics of high-wealth households*. Wealth Tax Commission. Available at: https://www.wealthandpolicy.com/wp/EP1_Distribution.pdf (Accessed: 10 July 2024).

Advani, A., Chamberlain, E. and Summers, A. (2020a) *A Wealth Tax for the UK*. Wealth Tax Commission. Available at: https://www.wealthandpolicy.com/wp/WealthTaxFinalReport.pdf (Accessed: 10 July 2024).

Advani, A., Hughson, H. and Tarrant, H. (2020c) *Revenue and distributional modelling for a wealth tax*. Wealth Tax Commission. Available at: https://doi.org/10.47445/113.

Advani, A. and Tarrant, H. (2021) 'Behavioural responses to a wealth tax', *Fiscal Studies*, 42(3–4), pp. 509–537. Available at: https://doi.org/10.1111/1475-5890.12283.

Alstadsaeter, A. *et al*. (2024) 'Global Tax Evasion Report 2024'. Eu-Tax Observatory. Available at: https://shs.hal.science/halshs-04563948 (Accessed: 27 August 2025).

Alvaredo, F., Garbinti, B. and Piketty, T. (2017) 'On the Share of Inheritance in Aggregate Wealth: Europe and the USA, 1900–2010', *Economica*, 84(334), pp. 239–260. Available at: https://doi.org/10.1111/ecca.12233.

Apostel, A. and O'Neill, D.W. (2022) 'A one-off wealth tax for Belgium: Revenue potential, distributional impact, and environmental effects', *Ecological Economics*, 196, p. 107385. Available at: https://doi.org/10.1016/j.ecolecon.2022.107385.

Bailey, D. (2015) 'The Environmental Paradox of the Welfare State: The Dynamics of Sustainability', *New Political Economy*, 20(6), pp. 793–811. Available at: https://doi.org/10.1080/13563467.2015.1079169.

Balata, F., Wright, H. and Sriskandarajah, D. (2025) 'Exploring an Extreme Wealth Line'. New Economics Foundation.

Bärnthaler, R. (2024) 'When enough is enough: Introducing sufficiency corridors to put techno-economism in its place', *Ambio*, 53(7), pp. 960–969. Available at: https://doi.org/10.1007/s13280-024-02027-2.

Bärnthaler, R. *et al*. (2026) 'Conceptualizing transformative climate action: insights from sufficiency research', *Climate Policy*, 0(0), pp. 1–20. Available at: https://doi.org/10.1080/14693062.2025.2494782.

Bärnthaler, R. and Gough, I. (2023) 'Provisioning for sufficiency: envisaging production corridors', *Sustainability: Science, Practice and Policy*, 19(1), p. 2218690. Available at: https://doi.org/10.1080/15487733.2023.2218690.

Bohnenberger, K. (2023) 'Peaks and gaps in eco-social policy and sustainable welfare: A systematic literature map of the research landscape', *European Journal of Social Security*, 25(4), pp. 328–346. Available at: https://doi.org/10.1177/13882627231214546.

Buch-Hansen, H. and Koch, M. (2019) 'Degrowth through income and wealth caps?', *Ecological Economics*, 160, pp. 264–271. Available at: https://doi.org/10.1016/j.ecolecon.2019.03.001.

Büchs, M. *et al.* (2024b) 'Emission inequality: Comparing the roles of income and wealth in Belgium and the United Kingdom', *Journal of Cleaner Production*, 467, p. 142818. Available at: https://doi.org/10.1016/j.jclepro.2024.142818.

Büchs, M. (2021b) 'Sustainable welfare: How do universal basic income and universal basic services compare?', *Ecological Economics*, 189, p. 107152. Available at: https://doi.org/10.1016/j.ecolecon.2021.107152.

Büchs, M., Ivanova, D. and Schnepf, S.V. (2021) 'Fairness, effectiveness, and needs satisfaction: new options for designing climate policies', *Environmental Research Letters*, 16(12), p. 124026. Available at: https://doi.org/10.1088/1748-9326/ac2cb1.

Büchs, M. and Koch, M. (2019) 'Challenges for the degrowth transition: The debate about wellbeing', *Futures*, 105, pp. 155–165. Available at: https://doi.org/10.1016/j.futures.2018.09.002.

Büchs, M., Koch, M. and Lee, J. (2024a) 'Sustainable Welfare: Decoupling Welfare from Growth and Prioritising Needs Satisfaction for All', *De Gruyter Handbook of Degrowth*. Berlin/Boston: Walter de Gruyter GmbH. Available at: http://ebookcentral.proquest.com/lib/leeds/detail.action?docID=31168140 (Accessed: 27 August 2025).

Cahen-Fourot, L. and Lavoie, M. (2016) 'Ecological monetary economics: A post-Keynesian critique', *Ecological Economics*, 126, pp. 163–168. Available at: https://doi.org/10.1016/j.ecolecon.2016.03.007.

Chakraborty, R. *et al.* (2019) 'Is the Top Tail of the Wealth Distribution the Missing Link between the Household Finance and Consumption Survey and National Accounts?', *Journal of Official Statistics*, 35(1), pp. 31–65. Available at: https://doi.org/10.2478/jos-2019-0003.

Chancel, L. *et al.* (2025) 'Climate Inequality Report 2025'.

Chancel, L. and Rehm, Y. (2025) 'Accounting for the carbon footprint of capital ownership advances the understanding of emission inequality', *Climatic Change*, 178(11), p. 211. Available at: https://doi.org/10.1007/s10584-025-04044-w.

Christensen, M.-B. *et al.* (2023) *Survival of the Richest: How we must tax the super-rich now to fight inequality*. Oxfam. Available at: https://doi.org/10.21201/2023.621477.

Dietz, R. and O'Neill, D.W. (2013) *Enough is enough: building a sustainable economy in a world of finite resources*. London: Routledge.

Disslbacher, F. *et al.* (2022) 'European Rich List Database (ERLDB)', 1. Available at: https://doi.org/10.17632/zpsr99hn35.1.

ECB (2020) 'Understanding household wealth: linking macro and micro data to produce distributional financial accounts'. Available at: https://www.ecb.europa.eu/pub/pdf/scpsps/ecb.sps37~433920127f.en.pdf (Accessed: 11 June 2025).

ECB (2024) 'Experimental Distributional Wealth Accounts (DWA) for the household sector Methodological note'. Available at: https://data.ecb.europa.eu/sites/default/files/2024-01/DWA%20Overview%20note_0.pdf (Accessed: 11 June 2025).

Engel, J. *et al.* (2022) *Developing reconciled quarterly distributional national wealth: insight into inequality and wealth structures*. Publications Office of the European Union. Available at: https://data.europa.eu/doi/10.2866/412495 (Accessed: 27 August 2025).

Engler, J.-O. *et al.* (2024) '15 years of degrowth research: A systematic review', *Ecological Economics*, 218, p. 108101. Available at: https://doi.org/10.1016/j.ecolecon.2023.108101.

European Environmental Bureau (2024) 'The case for a wealth tax'. Available at: https://eeb.org/library/the-case-for-a-wealth-tax/ (Accessed: 27 August 2025).

Fagereng, A. *et al.* (2018) 'Heterogeneity and Persistence in Returns to Wealth'. Available at: https://www.imf.org/en/Publications/WP/Issues/2018/07/27/Heterogeneity-and-Persistence-in-Returns-to-Wealth-46095 (Accessed: 27 August 2025).

Fanning, A.L., O'Neill, D.W. and Büchs, M. (2020) 'Provisioning systems for a good life within planetary boundaries', *Global Environmental Change*, 64, p. 102135. Available at: https://doi.org/10.1016/j.gloenvcha.2020.102135.

Fitzpatrick, N., Parrique, T. and Cosme, I. (2022) 'Exploring degrowth policy proposals: A systematic mapping with thematic synthesis', *Journal of Cleaner Production*, 365, p. 132764. Available at: https://doi.org/10.1016/j.jclepro.2022.132764.

Gore, T. (2022) 'Confronting carbon inequality'. Oxfam. Available at: https://www.oxfam.org/en/research/confronting-carbon-inequality (Accessed: 27 August 2025).

Gough, I. (2022) 'Two Scenarios for Sustainable Welfare: A Framework for an Eco-Social Contract', *Social Policy and Society*, 21(3), pp. 460–472. Available at: https://doi.org/10.1017/S1474746421000701.

Haberl, H. *et al.* (2020) 'A systematic review of the evidence on decoupling of GDP, resource use and GHG emissions, part II: synthesizing the insights', *Environmental Research Letters*, 15(6), p. 065003. Available at: https://doi.org/10.1088/1748-9326/ab842a.

Hickel, J. (2017) *The divide: a brief guide to global inequality and its solutions*. London: William Heinemann.

Hickel, J. *et al.* (2022) 'Degrowth can work — here's how science can help', *Nature*, 612(7940), pp. 400–403. Available at: https://doi.org/10.1038/d41586-022-04412-x.

Hirvilammi, T. and Koch, M. (2020) 'Sustainable Welfare beyond Growth', *Sustainability*, 12(5), p. 1824. Available at: https://doi.org/10.3390/su12051824.

Jackson, A. (2024) 'Wealth taxes and bonds'.

Jackson, T. and Jackson, A. (2021) *Beyond the Debt Controversy: Fiscal and monetary policy for the post-pandemic era*. All party parliamentary group on limits to growth. Available at: https://limits2growth.org.uk/wp-content/uploads/Beyond-the-Debt-Controversy-web.pdf (Accessed: 10 May 2025).

Jackson, T. and Victor, P.A. (2016) 'Does slow growth lead to rising inequality? Some theoretical reflections and numerical simulations', *Ecological Economics*, 121, pp. 206–219. Available at: https://doi.org/10.1016/j.ecolecon.2015.03.019.

Kakwani, N.C. (1977) 'Measurement of Tax Progressivity: An International Comparison', *The Economic Journal*, 87(345), pp. 71–80. Available at: https://doi.org/10.2307/2231833.

Kallis, G. *et al.* (2025) 'Post-growth: the science of wellbeing within planetary boundaries', *The Lancet Planetary Health*, 9(1), pp. e62–e78. Available at: https://doi.org/10.1016/S2542-5196(24)00310-3.

Kapeller, J., Leitch, S. and Wildauer, R. (2023) 'Can a European wealth tax close the green investment gap?', *Ecological Economics*, 209, p. 107849. Available at: https://doi.org/10.1016/j.ecolecon.2023.107849.

Kemp-Benedict, E. (2025) 'Transitioning to a sustainable economy: A preliminary degrowth macroeconomic model', *Ecological Economics*, 237, p. 108700. Available at: https://doi.org/10.1016/j.ecolecon.2025.108700.

Keynes, J.M. (1936) *The General Theory of Employment, Interest and Money*. Macmillan. Available at: https://www.perlego.com/book/4195823/the-general-theory-of-employment-interest-and-money-pdf (Accessed: 27 August 2025).

Khan, J. *et al.* (2023) 'Ecological ceiling and social floor: public support for eco-social policies in Sweden', *Sustainability Science*, 18(3), pp. 1519–1532. Available at: https://doi.org/10.1007/s11625-022-01221-z.

Knight, K.W., Schor, J.B. and Jorgenson, A.K. (2017) 'Wealth Inequality and Carbon Emissions in High-income Countries', *Social Currents*, 4(5), pp. 403–412. Available at: https://doi.org/10.1177/2329496517704872.

Koch, M. (2022) 'Social Policy Without Growth: Moving Towards Sustainable Welfare States', *Social Policy and Society*, 21(3), pp. 447–459. Available at: https://doi.org/10.1017/S1474746421000361.

Koch, M. and Hansen, A.R. (2024) 'Welfare within planetary limits: deep transformation requires holistic approaches'. Available at: https://doi.org/10.1332/TIZB1819.

Krenek, A. and Schratzenstaller, M. (2018) *A European net wealth tax*. Working Paper 561. WIFO Working Papers. Available at: https://www.econstor.eu/handle/10419/179315 (Accessed: 27 August 2025).

Landais, C., Saez, E. and Zucman, G. (2020) 'A progressive European wealth tax to fund the European COVID response', *VoxEU* [Preprint]. Available at: https://voxeu.org/ (Accessed: 6 March 2026).

Laruffa, F. (2025) 'Eco-social policies, capitalism and the horizon of emancipatory politics', *Critical Social Policy*, 45(2), pp. 259–279. Available at: https://doi.org/10.1177/02610183241262733.

Lawrence, M. *et al.* (2024) 'Global polycrisis: the causal mechanisms of crisis entanglement', *Global Sustainability*, 7. Available at: https://doi.org/10.1017/sus.2024.1.

Loutzenhiser, G. and Mann, E. (2020) *Liquidity issues: solutions for the asset rich, cash poor*. Wealth Tax Commission. Available at: https://www.wealthandpolicy.com/wp/EP10_Liquidity.pdf (Accessed: 21 July 2024).

Marti, S., Martínez, I.Z. and Scheuer, F. (2023) 'Does a progressive wealth tax reduce top wealth inequality? Evidence from Switzerland', *Oxford Review of Economic Policy*, 39(3), pp. 513–529. Available at: https://doi.org/10.1093/oxrep/grad025.

Mazzucato, M. (2018) *The value of everything: making and taking in the global economy*. London: Allen Lane, an imprint of Penguin Books.

Murphy, M.P. (2023) *Creating an Ecosocial Welfare Future: Making It Happen*. 1st edn. Policy Press. Available at: https://www.perlego.com/book/4141783/creating-an-ecosocial-welfare-future-making-it-happen-pdf (Accessed: 27 August 2025).

O'Donovan, N. (2021) 'One-off wealth taxes: theory and evidence', *Fiscal Studies*, 42(3–4), pp. 565–597. Available at: https://doi.org/10.1111/1475-5890.12277.

OECD (2018) *The Role and Design of Net Wealth Taxes in the OECD*. OECD (OECD Tax Policy Studies). Available at: https://doi.org/10.1787/9789264290303-en.

OECD (no date) *OECD Data Explorer • A comparative table of countries in the global database*. Available at: https://data-explorer.oecd.org/vis?fs[0]=Topic%2C1%7CTaxation%23TAX%23%7CGlobal%20tax%20revenues%23TAX_GTR%23&pg=0&fc=Topic&bp=true&snb=150&df[ds]=dsDisseminateFinalDMZ&df[id]=DSD_REV_COMP_GLOBAL%40DF_RSGLOBAL&df[ag]=OECD.CTP.TPS&dq=..S13._T..XDC.A&to[TIME_PERIOD]=false&pd=2017%2C2017&vw=tb (Accessed: 29 August 2025).

Olk, C., Schneider, C. and Hickel, J. (2023) 'How to pay for saving the world: Modern Monetary Theory for a degrowth transition', *Ecological Economics*, 214, p. 107968. Available at: https://doi.org/10.1016/j.ecolecon.2023.107968.

O'Neill, D.W. *et al.* (2018) 'A good life for all within planetary boundaries', *Nature Sustainability*, 1(2), pp. 88–95. Available at: https://doi.org/10.1038/s41893-018-0021-4.

Osier, G. (2016) *Unit non-response in household wealth surveys*. ECB Statistics Paper. Available at: https://www.ecb.europa.eu/pub/pdf/scpsps/ecbsp15.en.pdf (Accessed: 10 June 2025).

Parkes, H. and Johns, M. (2024) *Supporting the Status Quo*. Institute for Public Policy Research. Available at: Available from: https://ippr-org.files.svdcdn.com/production/Downloads/Supporting_the_status_quo_August24.pdf (Accessed: 21 October 2024).

Parrique, T. (2019) *The political economy of degrowth*. Université Clermont Auvergne. Available at: https://theses.hal.science/tel-02499463/file/2019CLFAD003_PARRIQUE.pdf (Accessed: 12 February 2025).

Piketty, T. (2014) *Capital in the Twenty-First Century*. Harvard University Press.

Piketty, T. and Saez, E. (2014) 'Inequality in the long run', 344, p. 838 to 843.

Piketty, T. and Zucman, G. (2014) 'Capital is Back: Wealth-Income Ratios in Rich Countries 1700–2010 *', *The Quarterly Journal of Economics*, 129(3), pp. 1255–1310. Available at: https://doi.org/10.1093/qje/qju018.

Raphael, R. *et al.* (2024) 'Postgrowth welfare systems: a view from the Nordic context'. Available at: https://doi.org/10.1332/27528499Y2024D000000026.

Robeyns, I. *et al.* (2021) 'How Rich is Too Rich? Measuring the Riches Line', *Social Indicators Research*, 154(1), pp. 115–143. Available at: https://doi.org/10.1007/s11205-020-02552-z.

Robeyns, I. (2022) 'Why Limitarianism?', *Journal of Political Philosophy*, 30(2), pp. 249–270. Available at: https://doi.org/10.1111/jopp.12275.

Rowlingson, K., Sood, A. and Tu, T. (2020) *Public attitudes to a wealth tax*. CAGE. Available at: https://doi.org/10.47445/102.

Saez, E. and Zucman, G. (2019) 'Progressive Wealth Taxation', *Brookings Papers on Economic Activity*, 2019(2), pp. 437–533.

Saez, E. and Zucman, G. (2022) 'Wealth Taxation: Lessons from History and Recent Developments', *AEA Papers and Proceedings*, 112, pp. 58–62. Available at: https://doi.org/10.1257/pandp.20221055.

Sakschewski, B. *et al.* (2025) *Planetary Health Check 2025*. Potsdam Institute for Climate Impact Research. Available at: https://www.planetaryhealthcheck.org/wp-content/uploads/PlanetaryHealthCheck2025.pdf (Accessed: 8 October 2025).

Schmiel, U. (2024) 'Wealth taxation of individuals and equity: A political-cultural market theory perspective', *Critical Perspectives on Accounting*, 99, p. 102465. Available at: https://doi.org/10.1016/j.cpa.2022.102465.

Stratford, B. (2020) 'The Threat of Rent Extraction in a Resource-constrained Future', *Ecological Economics*, 169, p. 106524. Available at: https://doi.org/10.1016/j.ecolecon.2019.106524.

Tian, P. *et al.* (2024) 'Keeping the global consumption within the planetary boundaries', *Nature*, 635(8039), pp. 625–630. Available at: https://doi.org/10.1038/s41586-024-08154-w.

Vadén, T. *et al.* (2020) 'Decoupling for ecological sustainability: A categorisation and review of research literature', *Environmental Science & Policy*, 112, pp. 236–244. Available at: https://doi.org/10.1016/j.envsci.2020.06.016.

Varoufakis, Y. (2024) *Technofeudalism*. Available at: https://www.penguin.co.uk/books/451795/technofeudalism-by-varoufakis-yanis/9781529926095 (Accessed: 9 March 2026).

Vogel, J. and Hickel, J. (2023) 'Is green growth happening? An empirical analysis of achieved versus Paris-compliant CO2–GDP decoupling in high-income countries', *The Lancet Planetary Health*, 7(9), pp. e759–e769. Available at: https://doi.org/10.1016/S2542-5196(23)00174-2.

Waltl, S.R. (2022) 'Wealth Inequality: A Hybrid Approach Toward Multidimensional Distributional National Accounts In Europe', *Review of Income and Wealth*, 68(1), pp. 74–108. Available at: https://doi.org/10.1111/roiw.12519.